\documentclass[11pt,a4paper]{article}

\usepackage{authblk}
\usepackage{tabularx,booktabs}
\usepackage{fullpage}
\usepackage{graphicx}
\usepackage{lipsum}
\usepackage[utf8]{inputenc}
\usepackage[margin=1.0in]{geometry}
\usepackage[hidelinks]{hyperref}
\usepackage[T1]{fontenc}
\usepackage[utf8]{inputenc}
\usepackage{amsmath,amssymb}
\usepackage{bm}
\usepackage{placeins}
\usepackage{dsfont}
\usepackage{url}
\usepackage{svg}
\usepackage{graphics}
\usepackage[export]{adjustbox}
\usepackage{todonotes}
\usepackage[labelfont=bf]{caption}
\usepackage{subcaption}
\usepackage{soul}
\usepackage{bbding}
\usepackage{dsfont}
\usepackage[square,numbers,sort&compress]{natbib} 

\usepackage{graphicx}   
\usepackage{booktabs}   
\usepackage{array}      
\usepackage{longtable}  

\newif\ifblind
\blindfalse

\hypersetup{
  pdfauthor={Anonymous},
  pdftitle={Near-Lossless Model Compression Enables Longer Context Inference in DNA Large Language Models}
}

\usepackage{amsmath,amssymb}

\newcommand{\junk}[1]{}

\begin{document}
\thispagestyle{empty}
\def\ourmethod{\textsc{Focus}}

\captionsetup[subfigure]{justification=raggedright,singlelinecheck=false}

\ifblind
  \title{Near-Lossless Model Compression Enables Longer Context Inference in DNA Large Language Models}
  \author[ ]{\textbf{Anonymous Authors}}
  \affil[ ]{\small Under double-blind review}
  \date{}  
\else
  \title{Near-Lossless Model Compression Enables Longer Context Inference in DNA Large Language Models}

  \setcounter{Maxaffil}{2}
  \author[1]{Rui Zhu}
  \author[2]{Xiaopu Zhou}
  \author[3]{Haixu Tang}
  \author[2]{Stephen W. Scherer}
  \author[1,\thanks{Corresponding author.}]{Lucila Ohno-Machado}

  \affil[1]{\small Yale School of Medicine, New Haven, USA}
  \affil[2]{\small The Hospital for Sick Children, Toronto, Canada}
  \affil[3]{\small Luddy School of Informatics, Computing, and Engineering, Indiana University Bloomington, USA}
  \date{}
\fi

\maketitle

\vspace{-30 pt}
\begin{abstract}

Trained on massive cross-species DNA corpora, DNA Large language Models (LLMs) learned the fundamental ``grammar'' and evolutionary patterns of genomic sequences. This makes them powerful priors for DNA sequence modeling, particularly across long distances. Yet, two major constraints hinder their use: the quadratic computational cost of self-attention and the expanding memory required for key-value ($\mathbf{KV}$) caches during autoregressive decoding. These constraints force the use of heuristics like fixed-window truncation or sliding windows, which compromise fidelity on ultra-long sequences by discarding distant information. 
We introduce \ourmethod\ (\textbf{F}eature-\textbf{O}riented \textbf{C}ompression for \textbf{U}ltra-long \textbf{S}elf-attention), a progressive context-compression module that can be plugged into pretrained DNA LLMs. \ourmethod\ combines the established $k$-mer representation in genomics with learnable hierarchical compression: it inserts \textit{summary tokens} at $k$-mer granularity and progressively compresses attention key/value activations across multiple Transformer layers, retaining only the summary $\mathbf{KV}$ states across windows while discarding ordinary-token $\mathbf{KV}$. A shared-boundary windowing scheme yields a stationary cross-window interface that propagates long-range information with minimal loss.
We validate \ourmethod\ on an Evo-2--based DNA LLM, which was fine-tuned on GRCh38 chromosome~1 with self-supervised training and randomized compression schedules to promote robustness across compression ratios. On held-out human chromosomes,  \ourmethod\ demonstrated near-lossless fidelity: compressing a 1\,kb context into only 10 summary tokens ($\sim$100$\times$) shifted average per-nucleotide
probability by only $\sim4\times10^{-4}$. Compared to the baseline, \ourmethod\ reduces $\mathbf{KV}$-cache memory and converts effective inference scaling from $O(N^2)$ to near-linear $O(N)$, enabling $\sim$100$\times$ longer inference windows on commodity GPUs with near-lossless fidelity.

\end{abstract}

\newpage

\section{Introduction}
  
\label{sec:intro}
Modeling long-range genomic dependencies is central to modern bioinformatics. Classical statistical approaches---such as $k$-mer counting and motif scanning, hidden Markov models, and shallow discriminative models---have provided valuable local signals but are typically constrained by limited labeled data, strong independence or locality assumptions, and difficulty scaling to megabase (Mb) contexts \citep{Grant2011FIMO, Eddy1998HMM, Ghandi2014gkmSVM, Libbrecht2015ML}. By contrast, DNA large language models (DNA LLMs) trained on massive multi-species corpora can internalize the ``grammar'' of the genome (e.g., motif co-occurrence, evolutionary regularities, and distal regulatory couplings), offering strong priors for sequence modeling at scale. Yet, in practice, even DNA LLMs struggle with \emph{ultra}-long sequences. Their Transformer the backbone incurs a quadratic $O(N^2)$ self-attention cost and maintain a growing memory footprint through the \textit{key--value ($\mathbf{KV}$) cache}, i.e., the per-layer storage of historical keys ($\mathbf{K}$) and values ($\mathbf{V}$) needed for autoregressive decoding \citep{Zaheer2020BigBird, Rae2020Compressive}. Consequently, standard inference pipelines truncate long sequence context to a fixed window or use sliding-window heuristics. These methods inevitably discard distal history \citep{Dai2019TransXL}, precisely where much of the biological information, such as enhancer--promoter links, TAD (Topologically Associating Domain) boundary effects, replication timing, repeats, and structural variations (SVs) spanning tens of kilobases to megabases, resides \citep{Dixon2012TAD, Marchal2019Replication, Sedlazeck2018LongSV, Treangen2012Repeats}. 

To reconcile the need for Mb-scale evidence with commodity hardware limits, we leverage a fundamental genomics abstraction: the \emph{$k$-mer}, i.e., a contiguous stretch of $k$ nucleotides, This concept is the building block of the \textit{de Bruijn graph}, an essential data structure for sequence analysis tasks such as indexing ~\cite{marchet2021data,shiryev2024indexing}, fragment assembly ~\cite{pevzner2001eulerian,chikhi2014informed}, and abundance estimation ~\cite{melsted2014kmerstream,benoit2016multiple}.  Building on this idea, we introduce \ourmethod (\textbf{F}eature-\textbf{O}riented \textbf{C}ompression for \textbf{U}ltra-long \textbf{S}elf-attention): a learnable, hierarchical context-compression mechanism. By combining $k$-mer--aware summarization with modern LLM compression, \ourmethod\ plugs into pretrained DNA LLMs and compactly summarizes long DNA sequences with minimal loss.

\begin{figure}[htbp]
  \centering
  \includegraphics[width=0.8\textwidth]{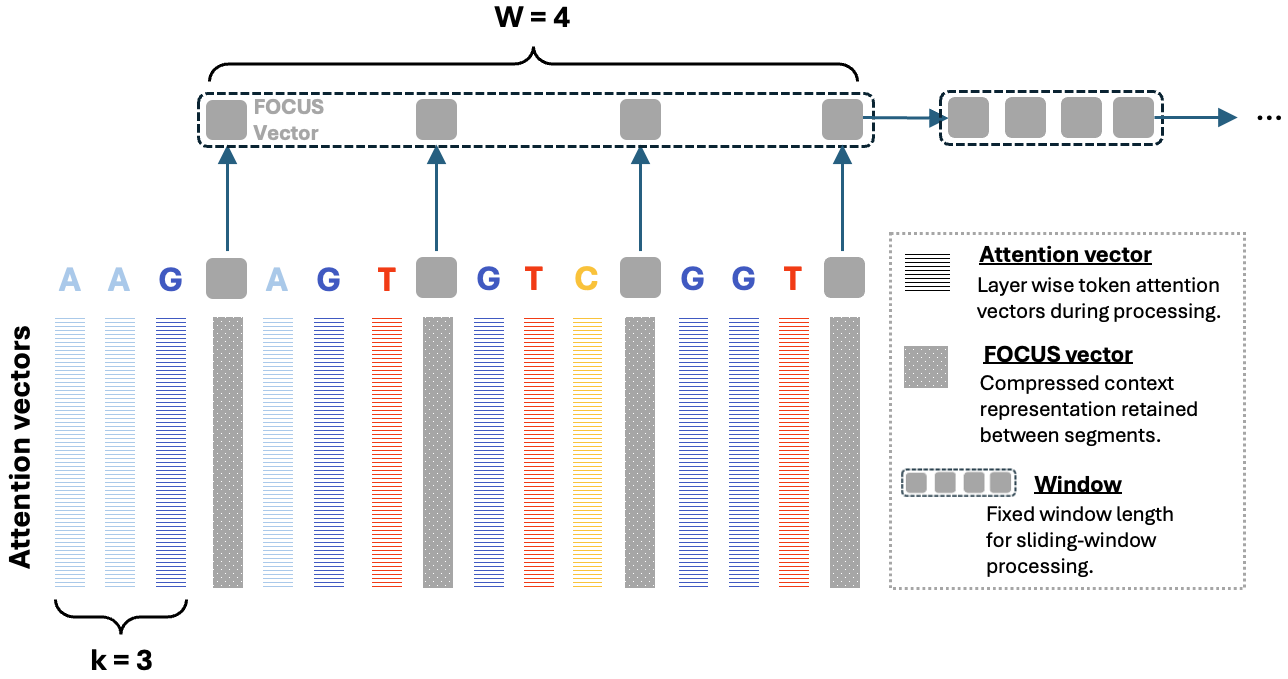}
  \caption{\textbf{\ourmethod{} at $k$-mer granularity with sliding windows.}
Illustrative example with $k{=}3$ and window size $W{=}4$.
After every $k$ ordinary bases, a learnable \ourmethod\ token (gray square) is inserted and,
through a dedicated attention module, summarizes the immediately preceding $k$-mer into a compact
\emph{\ourmethod\ vector}. Generation proceeds in fixed windows ($W$): only \ourmethod\ states are
retained and carried across windows, while ordinary-token states are not kept. Legend:
layerwise attention vectors (striped), \ourmethod\ vector (dotted gray), window (outlined blocks).}
  \label{fig:Fig_1} 
\end{figure}

Concretely, as shown in ~\autoref{fig:Fig_1}, \ourmethod\ inserts learnable summary tokens at the $k$-mer level, retaining only their $\mathbf{KV}$ states states while discarding those from all ordinary base tokens. A shared-boundary windowing scheme creates a stationary interface, allowing long-range information to propagate hop-by-hop. Consequently, the number of retained states scales with the number of summary tokens, not the original sequence length. This approach reduces $\mathbf{KV}$-cache memory by roughly two orders of magnitude and converts inference scaling from $O(N^2)$ to near-linear $O(N)$, where $N$ is the context length, enabling $\sim\!100\times$ longer inference windows on commodity GPUs with near-lossless fidelity.

We instantiate \ourmethod\ on an Evo-2-7b DNA LLM, training only the compression parameters on the GRCh38 chromosome 1 with a self-supervised objective and randomized compression schedules. Our results demonstrate near-lossless fidelity on held-out human chromosomes and other corpora. \autoref{sec:Result} highlights more findings: (i) fidelity degrades predictably with increasing sequence length due to cross-window accumulation, and (ii) this discrepancy is substantially reduced by enlarging the window or using a finer granularity. In terms of scalability, we show that peak memory remains essentially flat with context, and single-GPU inference scales to $\sim\!80$k tokens with almost constant memory usage.

\ourmethod\ is architecture-agnostic and label-free method that can be dropped into any pretrained DNA LLMs to unlock genome-scale context while preserving accuracy. The approach introduces explicit knobs ($W$ and $k$) to trade fidelity for throughput, though this means minor approximation errors can accumulate across many windows. 
\ourmethod\ opens practical routes to Mb-scale reasoning for tasks that intrinsically require distal evidence, such as SV breakpoint interpretation, repeat disambiguation, and long-range regulatory inference. 
Future extensions of the method could include multi-resolution summaries, adaptive $k$, coupling with efficient attention or retrieval, and broader whole-genome task evaluations.

\vspace{-10 pt}
\paragraph{Contributions.}
(i) A $k$-mer--aware, learnable hierarchical compression module for DNA LLMs that retains only summary $\mathbf{KV}$ states and uses shared-boundary windows for stable, ultra-long propagation; (ii) a training recipe that requires no architectural changes or labeled data and is robust across compression ratios; (iii) extensive evaluations showing $\sim$100$\times$ longer effective context with near-lossless fidelity and near-linear memory scaling.

\vspace{-10 pt}
\section{Background}
  
\label{sec:background}

\vspace{-5 pt}
\subsection{Large Language Models}
\label{subsec:Large Language Models}
Transformer LLMs compute self-attention by projecting each token into query ($\mathbf{Q}$), key ($\mathbf{K}$), and value ($\mathbf{V}$) vectors. At each position, attention weights are obtained by matching the current $\mathbf{Q}$ against all past $\mathbf{K}$'s and then aggregating the corresponding $\mathbf{V}$'s to form a contextualized representation. During autoregressive inference, implementations cache all historical $K$ and $V$ tensors at every layer---the \emph{$\mathbf{KV}$ cache}---so that each new token only computes its own $Q$ and attends to the stored history without recomputing earlier states. While this cache greatly reduces redundant compute, its memory usage grows linearly with context length and scales with layers, heads, and hidden dimensions; meanwhile, attention itself incurs quadratic time/space with sequence length. Consequently, very long contexts become impractical, and practitioners fall back to fixed-window truncation or sliding windows that discard distal evidence. Our method specifically targets this bottleneck by learning to preserve only a small set of information-critical $\mathbf{KV}$ states while safely discarding redundant ones.

\vspace{-5 pt}
\subsection{DNA Large Language Models}
\label{subsec:DNA LLM}
DNA large language models (LLMs) treat genomes as sequences over a four-letter alphabet and learn statistical regularities that span from local motifs to long-range dependencies. Early work established the paradigm: DNABERT\cite{Ji2021DNABERT} adapted the BERT architecture to $k$-mer tokenization and showed that a single pre-trained model can transfer to promoter, splice-site, and TF-binding prediction with minimal supervision. GenSLM\citep{Zvyagin2023GenSLM} scaled foundation modeling to whole viral genomes, pretraining on $>$100M prokaryotic genes and fine-tuning on 1.5M SARS-CoV-2 genomes to track evolutionary dynamics and variants of concern. These efforts demonstrated that transformer-style models learn biologically meaningful representations, but their context windows typically remained at kilobase scale.

Recent models push both scale and context and now define the state of the art. Evo~\cite{Nguyen2024Evo} is a 7B-parameter autoregressive genomic foundation model built on the StripedHyena long-context architecture. Trained at single-nucleotide resolution on $\sim$2.7M prokaryotic and phage genomes, Evo supports $\sim$131\,kb context and generalizes across DNA, RNA, and proteins \citep{Nguyen2024Evo, Marchal2024EvoHighlight}. Beyond zero-shot functional inference (e.g., gene essentiality and regulatory activity), Evo can \emph{generate} long, genome-scale sequences and even co-design multi-component systems such as CRISPR protein--RNA complexes \citep{Nguyen2024Evo}.

Building on this, Evo-2~\cite{Brixi2025Evo2} increases both data and context: trained on $\sim$9.3T nucleotides from $>$128{,}000 genomes spanning all domains of life, and using StripedHyena 2, it extends usable context to $\sim$1\,Mb at single-base resolution \citep{Brixi2025Evo2}. Evo-2 reports strong zero-shot and fine-tuned performance on variant-effect prediction (e.g., \textit{BRCA1}) and enables design of long synthetic sequences resembling natural genomes. Together, Evo and Evo-2 demonstrate that long-context DNA LLMs can reason over distal regulatory signals and structural variation at scales that were previously impractical. 
However, even with aggressive systems optimizations (e.g., mixed precision and kernel fusion) to reduce cache pressure, the compute and memory footprint remains substantial: in a highly optimized BioNeMo implementation of Evo-2 7B 
requires at least eight H100 GPUs to perform inference on a 1M-base sequence\cite{Brixi2025Evo2} (an $8\times$H100 server typically costs \$250k--\$350k in 2025\cite{supermicro_sys821ge_store_2025,thinkmate_sys821ge_2025,uvation_xe9680_h100_2025,arccompute_xe9680_starting_2025,supermicro_h100_pcie_price_2025}, such systems beyond the reach of most academic labs). 
Consequently, many genomics groups can realistically deploy such models only for short-range analyses\cite{zhu2025advancing} (e.g., single-variant and haplotype inference), whereas Mb-scale applications such as structural-variation detection or whole-genome inference remain impractical.

\vspace{-5 pt}
\subsection{Motivation \& Challenge}
\label{subsec:Motivation}
\paragraph{Why truly long context is essential for DNA LLMs.}
Recent advances in DNA large language models (LLMs) have extended contextual inference to approximately 1 Mb, allowing the capture of exon--intron boundaries and the regulatory landscape within a single gene body. However, this context length remains insufficient for modeling broader genomic dependencies, such as long-range linkage disequilibrium (LD) patterns observed in genome-wide association studies (GWAS) and trans-regulatory interactions mediated by chromatin looping across multiple topologically associating domains (TADs), which can span several megabases. These higher-order regulatory architectures govern how the same DNA sequence gives rise to diverse cellular and tissue identities, encompassing processes fundamental to normal physiology and disease pathogenesis, yet they remain beyond the modeling capacity of current DNA LLMs. Specifically, (i) Regulatory and 3D genomic signals span long distances. Enhancer--promoter interactions, insulator effects at TAD boundaries, and replication-timing domains often require $\mathrm{kb}$--$\mathrm{Mb}$ of joint context to interpret correctly \citep{Dixon2012TAD, Marchal2019Replication}. (ii) Structural variation (SV) is inherently long-range. Deletions, insertions, inversions, translocations, and copy-number variations create breakpoints whose interpretation depends on distal elements and repeat-rich neighborhoods; robust SV calling and impact assessment benefit from kilobases to megabases of flanking regions to distinguish true events from alignment artifacts and to deduce disruption of regulatory mechanisms \citep{Sedlazeck2018LongSV}. (iii) Repetitive and segmentally duplicated regions demand global disambiguation. Local signals alone are insufficient to resolve ambiguity without broader genomic context \citep{Treangen2012Repeats}. These use-cases all demand substantially longer effective context than standard attention windows can provide.

\vspace{-10 pt}
\paragraph{Where the bottlenecks lie: attention and the $\mathbf{KV}$ cache.}
Self-attention scales quadratically with sequence length, and the $\mathbf{KV}$ cache---i.e., the per-layer, per-head storage of all historical keys ($\mathbf{K}$) and values ($\mathbf{V}$) that future queries ($\mathbf{Q}$) must attend to---grows linearly with context. In long-context decoding, memory therefore scales with the number of historical tokens times the dimensionality, multiplied across layers and heads; beyond a few thousand tokens, this becomes the dominant constraint on commodity GPUs. Sliding-window approximations mitigate cost but irrevocably discard distal evidence, which is precisely what many genomic tasks require.

\vspace{-10 pt}
\paragraph{Our approach.}
\ourmethod\ introduces learnable, $k$-mer--aware summary tokens and retains only their $\mathbf{KV}$ states across windows while discarding ordinary $\mathbf{KV}$, so the number of persistent states grows with $L/k$ instead of $L$. A shared-boundary windowing scheme ensures information flows stably across windows with a stationary interface, yielding near-linear scaling in context length and large reductions in peak memory. 
In our experiments, compressing 1\,kb into $\sim$10 summary tokens induces only $\approx$0.2--0.4\% changes in per-token distributions across in-domain and out-of-domain benchmarks, while $\mathbf{KV}$ memory drops by $\sim$100$\times$ and effective inference scaling approaches $O(N)$.

\vspace{-10 pt}
\section{Methods}
  \vspace{-3 pt}

\subsection{Problem Definition and Method Intuition}
\label{subsec:Problem Definition}

\paragraph{Problem Definition.}
Let \( \Sigma=\{\mathrm{A},\mathrm{C},\mathrm{G},\mathrm{T}\} \) and \(x_{1:L}\in\Sigma^{L}\) be a DNA sequence.
Consider a pretrained autoregressive DNA language model \(M_{\theta}\) with maximum context length \(N\).
For any prefix \(x_{1:t}\) with \(t\le N\), the model computes the next--base distribution
\[
P_{\theta}\!\left(x_{t+1}\mid x_{1:t}\right).
\]
When \(L>N\), standard practice truncates the context to the most recent \(N\) bases (or uses a sliding window), which introduces approximation error.
Our goal is to endow \(M_{\theta}\) with a \emph{compression function} \(f\) such that, for \(L>N\),
the long context can be summarized into an effective representation whose size is on the order of \(N\) (or smaller), while preserving predictive fidelity.

Formally, for \(t>N\), let the remote history be \(x_{1:t-N}\) and define a compressed state
\[
c_t \;=\; f_{\phi}\!\left(x_{1:t-N}\right), \qquad \text{with effective length } \tilde N \lesssim N .
\]
The next--base distribution is then computed by conditioning on both the compressed state and the recent, uncompressed window:
\[
\tilde P_{\theta,\phi}\!\left(x_{t+1}\mid c_t,\; x_{t-N+1:t}\right) \;\approx\; P_{\theta}\!\left(x_{t+1}\mid x_{1:t}\right).
\]
We require (i) \emph{fidelity}: the approximation is close to the (ideal) full-context output, e.g.
\[
\mathbb{E}_{x\sim\mathcal D}\!\left[\mathrm{KL}\!\left(
P_{\theta}(\cdot\mid x_{1:t})
\;\big\|\;
\tilde P_{\theta,\phi}(\cdot\mid c_t, x_{t-N+1:t})
\right)\right] \text{ is small,}
\]
and (ii) \emph{efficiency}: computing \(c_t\) (preferably incrementally) and evaluating \(\tilde P_{\theta,\phi}\) should not exceed the time/memory complexity of running \(M_{\theta}\) on a length-\(N\) input.

To avoid ambiguity between the two uses of ``K,'' we write $k$ (lower case) for the $k$-mer length (i.e., the number of bases per segment), and reserve boldface $\mathbf{K}$ (together with $\mathbf{Q}$ and $\mathbf{V}$) for the attention \emph{Key}/\emph{Query}/\emph{Value} tensors. 
Accordingly, the \emph{$\mathbf{KV}$ cache} denotes the stored pair $(\mathbf{K}, \mathbf{V})$ for all past positions.

\paragraph{Method Intuition.}

In genomics, a \emph{$k$-mer} is a contiguous segment of $k$ nucleotides that captures local motifs and short-range syntax. 
As sketched in \autoref{fig:Fig_1}, we operate at this $k$-mer granularity: after every $k$ real bases we \emph{insert} a learnable \ourmethod\ token that uses a dedicated attention projection to \emph{aggregate} the salient information from the immediately preceding $k$-mer into a compact \emph{\ourmethod\ vector}. 
Downstream layers then \emph{retain only the Key/Value ($\mathbf{KV}$) states} of these \ourmethod\ tokens---optionally plus a tiny boundary tail---while discarding $\mathbf{KV}$ from ordinary tokens. 
During generation, the sequence is processed in fixed windows of size $W$; at each window boundary, a new \ourmethod\ token is produced, its $\mathbf{KV}$ is carried forward, and the window slides to the next segment (see the dashed window outline and gray squares in \autoref{fig:Fig_1}). 
Consequently, the amount of $\mathbf{KV}$ we preserve grows with the number of \ourmethod\ tokens rather than the raw sequence length, yielding memory that scales roughly with $L/k$ instead of $L$, with $k$ acting as the primary knob to trade memory/speed against fidelity (larger $k$ gives stronger compression but may increase approximation error). 
For completeness, the full mathematical formalism and training/inference details are provided in the \autoref{sec:Formal Methodology}.

\subsection{\ourmethod: Feature-Oriented Compression for Ultra-long Self-attention}
\label{subsec:Feature-Oriented Compression for Ultra-long Self-attention}

\paragraph{Setup and tokens.}
Given a DNA sequence $x_{1:L}$, we insert one \ourmethod{} token after every $k$ consecutive \emph{base} tokens (a $k$-mer). Let $M=\lceil L/k\rceil$ be the number of $k$-mers, and denote the $i$-th \ourmethod{} token by $S_i$ with its associated $k$-mer (the $k$ bases immediately to its left) denoted as $\mathcal{B}_i$. At designated layers, $S_i$ aggregates the salient information from $\mathcal{B}_i$ (and earlier summary states), thereby acting as a compact, trainable summary of the respective $k$-mer. All attention is strictly causal in token order.

\paragraph{Windowing with a shared boundary.}
We group \ourmethod{} tokens into \emph{windows} of fixed size $W$ in the \ourmethod{}-token domain (not base tokens). The $m$-th window contains an ordered block
\[
\big(S_{m,1}, S_{m,2}, \dots, S_{m,W}\big),
\]
and adjacent windows \emph{share one boundary token}:
\[
S_{m,1}\;\equiv\;S_{m-1,W}\qquad (m>1).
\]
Intuitively, each window carries forward the last summary from the previous window as its first element, so the model only needs to learn a \emph{single, canonical interaction pattern} among the $W$ \ourmethod{} tokens in a window and how they connect to earlier context; long-range propagation then follows by repeatedly applying the same pattern window-by-window. If $M$ is the total number of \ourmethod{} tokens, the number of windows is on the order of $M/W\approx L/(k\,W)$ (the exact count with a one-token overlap is $\lceil(M-1)/(W-1)\rceil$, which is asymptotically equivalent for large $M$).

\paragraph{Visibility and summarization rule.}
Within a window, base tokens interact locally while \ourmethod{} tokens carry the global channel:
\begin{itemize}
    \item A base token belongs to exactly one $k$-mer and only attends to past tokens within its own $k$-mer and to \emph{preceding} \ourmethod{} tokens in the same window (including the carried boundary $S_{m,1}$).
    \item The \ourmethod{} token $S_{m,j}$ summarizes its left $k$-mer $\mathcal{B}_{m,j}$ and may attend to the preceding \ourmethod{} tokens $\{S_{m,1},\ldots,S_{m,j-1}\}$ inside the same window (again including the carried boundary). Earlier base tokens outside the current window are \emph{not} visible; their information is only accessible through earlier \ourmethod{} tokens.
\end{itemize}
Within each window, we update every \ourmethod{} token by letting it look back only at two sources of information: (i) the $k$ base tokens that form its own segment, and (ii) the earlier \ourmethod{} tokens inside the same window (including the boundary token carried over from the previous window). It does \emph{not} access base tokens from previous windows and never looks ahead. The combination is implemented by a small causal-attention block in which the \ourmethod{} token queries this visible context and produces a single updated state that replaces its current representation. A complete, equation-level specification (including masks and parameterization) is provided in \autoref{sec:Formal Methodology}.

Intuitively, each \ourmethod{} token compresses its own $k$-mer and stitches that summary to the running context represented by earlier \ourmethod{} tokens. Because consecutive windows share their boundary token, the model only needs to learn a single within-window interaction pattern; applying the same pattern window-by-window propagates information across the full sequence while keeping memory proportional to the number of \ourmethod{} tokens.

\paragraph{Trainable summary module.}
Each \ourmethod{} token is backed by a lightweight, trainable module that operates alongside the frozen DNA LLM decoder. First, we learn a dedicated embedding vector for the special summary token: if the base hidden size is $d$, this adds exactly one $d$-dimensional parameter row that encodes how a fresh summary is inserted after every $k$ bases. Second, we attach a compact causal-attention adapter of size $\mathcal{O}(d^2)$ whose query/key/value projections act only on the visible context of a summary token---namely its own $k$-mer $\mathcal{B}{m,j}$ and the preceding summaries ${S_{m,1},\ldots,S_{m,j-1}}$ inside the same window. This adapter replaces the summary's hidden state by attending to those $k$ base features plus earlier summaries, producing a single condensed representation that will be retained in the $\mathbf{KV}$ cache. All other parameters of the DNA LLM backbone remain fixed; thus the total number of trainable weights is dominated by (i) the single summary-token embedding and (ii) the projection matrices of this window-local attention block, which collectively amount to a few tens of millions of parameters---tiny compared with the base model but sufficient to learn how information flows from each $k$-mer into its summary.

\paragraph{$\mathbf{KV}$ retention and memory profile.}
After completing a window, we \emph{discard} the key/value ($\mathbf{KV}$) states of its base tokens and \emph{retain only} the $\mathbf{KV}$ states of its $W$ \ourmethod{} tokens (the last of which is reused as the first token in the next window). Over the whole sequence, the number of retained summaries is approximately $M\approx L/k$. The effective $\mathbf{KV}$ compression ratio is therefore
\[
\gamma(L)\;=\;\frac{R_{\mathbf{KV}}^{\mathrm{kept}}(L)}{L}\;\approx\;\frac{1}{k}\qquad\text{for }L\gg k,
\]
while the \emph{active} memory during window processing is bounded by the $W$ \ourmethod{} tokens plus the current window's $k$-mer bases (a small, fixed multiple of $k$). In practice, compute scales with the window size (roughly $\mathcal{O}((Wk)^2)$ locally), and long-range state scales linearly with $L/k$.

\paragraph{Why the shared-boundary window helps.}
The boundary reuse $S_{m,1}\equiv S_{m-1,W}$ makes the cross-window interface \emph{stationary}: the model repeatedly applies the same within-window association pattern, needing only to learn how $W$ \ourmethod{} tokens (plus their local bases) interact under causality. Information from distant context is propagated hop-by-hop through these shared boundaries, yielding strong compression (memory $\propto L/k$) while preserving fidelity via learned, window-local summaries.
\subsection{Training Objective}
\label{subsec:training-objective}

We train \ourmethod{} using a standard cross-entropy language modeling objective, adapted to the segmented DNA sequence. At each position in the augmented sequence (including both base tokens and inserted summary tokens), the model predicts the next token. Let $y_1, y_2, \dots, y_{N'}$ denote the tokens in the output sequence (where $N' = N + M$ includes the inserted summary tokens). The training loss is:
\begin{equation}
\mathcal{L} = -\frac{1}{N'} \sum_{t=1}^{N'} \log P(y_{t} \mid y_{<t})\,,
\label{eq:crossentropy}
\end{equation}
where $P(y_t \mid y_{<t})$ is the model's predicted probability for the token $y_{t}$ at position $t$ given all previous tokens $y_{<t}$ (subject to the attention constraints described above). This objective encourages the model to accurately predict both the real nucleotides and the special summary tokens in their appropriate positions. During generation or downstream evaluations, the summary tokens can be omitted or treated as internal markers, while the base tokens produce the actual DNA sequence. By optimizing the loss in Eq.~\eqref{eq:crossentropy}, \ourmethod\ learns to balance local detail and global context: it must use the summarization tokens to carry information forward in order to predict distant bases correctly, thereby effectively learning the compression mechanism as part of the language modeling task.
  
\vspace{-10 pt}
\section{Results}
  \label{sec:Result}

\subsection{Experimental Setup: Models and Data}
\label{subsec:setup}
\textbf{Training corpus and distribution splits.} Evo-2 is pre-trained on \textit{OpenGenome2}, a large-scale compendium of genomes spanning all domains of life. To align our evaluation with this training distribution, we partition all test sequences into two groups: \emph{In-Distribution (ID)}, drawn from species represented in OpenGenome2, and \emph{Out-of-Distribution (OOD)}, drawn from species absent in OpenGenome2. Specifically, we use sequences from the \emph{MSL39} viral dataset for OOD evaluation, whose species are absent from OpenGenome2. Note that OpenGenome2 does not include viral genomes in its training data. We report results separately for these two groups throughout.

We compare two models: the uncompressed \textbf{Evo-2 7B} and the \textbf{\ourmethod{}-compressed Evo-2 7B} (abbrev. \ourmethod{}--Evo-2 7B).
\ourmethod{} uses a fixed window size $W{=}1024$; after every $k{=}100$ real bases, a learnable \ourmethod{} token is inserted and only the Key/Value ($\mathbf{KV}$) states of these tokens are retained across windows.
All \ourmethod{} parameters are trained on GRCh38 Chromosome~1; the remaining Evo-2 weights are kept frozen.
Unless otherwise stated, inputs are non-overlapping $1024$\,bp segments.
\emph{Within the ID group}, we further stratify evaluation into (i) held-out human genomic sequences from GRCh38 chromosomes other than Chromosome 1, and (ii) non-human genomic sequences that are included in OpenGenome2.
This stratification distinguishes generalization beyond the trained human chromosome from generalization across species within the pre-training distribution.

For details of the training procedure and hyperparameters, please refer to \autoref{sec:FOCUS hyper-parameters}.

\vspace{-4 pt}
\subsection{Evaluation Metrics}
\label{subsec:metrics}

We assess \ourmethod{} along two complementary axes: 
\emph{Fidelity}: how closely the compressed model reproduces the outputs of the original model on the same inputs---and \emph{Scalability}: how compression changes the GPU memory required to process long contexts.

For fidelity, we evaluate how closely the output of the compressed model matches the that of the original model on the same sequences. 
Unless otherwise noted, we take 1\,kb segments and run \emph{both} the uncompressed  Evo-2 7B (baseline) and the \ourmethod{}--Evo-2 7B on each segment as input. 
At every position in a segment, we record the model's \emph{next-base probability distribution} (i.e., the probabilities it assigns to the next token in the vocabulary). 
This yields, for each position, a pair of distributions---one from the baseline and one from the compressed model---that we can compare directly.

We quantify distributional differences between the output of the baseline and \ourmethod{} using five complementary metrics, listed here without formulas for readability:
\begin{enumerate}
  \item \textbf{L1 distance} --- sum of absolute differences between the two probability distributions.
  \item \textbf{L2 distance} --- root-mean-square difference, emphasizing larger deviations.
  \item \textbf{Hellinger distance} --- a symmetric, bounded measure that highlights overlap of probability mass.
  \item \textbf{Jensen--Shannon (JS) divergence} --- a symmetric, smoothed variant of KL that remains finite even when supports do not overlap.
  \item \textbf{KL divergence} --- a directional measure that penalizes probability mass missing from \ourmethod{} relative to the baseline. In here, we use the distribution from uncompressed model as the base.
\end{enumerate}
Formal definitions, estimation details, and implementation notes (e.g., any probability clipping and re-normalization) are provided in \autoref{sec:Metrics}.

For numerical stability we apply probability clipping at $10^{-12}$ and re-normalization.
A sequence-level score can also be defined as $\overline{D}(x)=\frac{1}{T}\sum_{t=1}^{T} D(\mathbf{p}_t,\mathbf{q}_t)$; the figures visualize the overall distribution of per-position metrics across all $(t,x)$ together with the median and IQR.

\begin{figure}[htbp]
  \centering
  \includegraphics[width=1\textwidth]{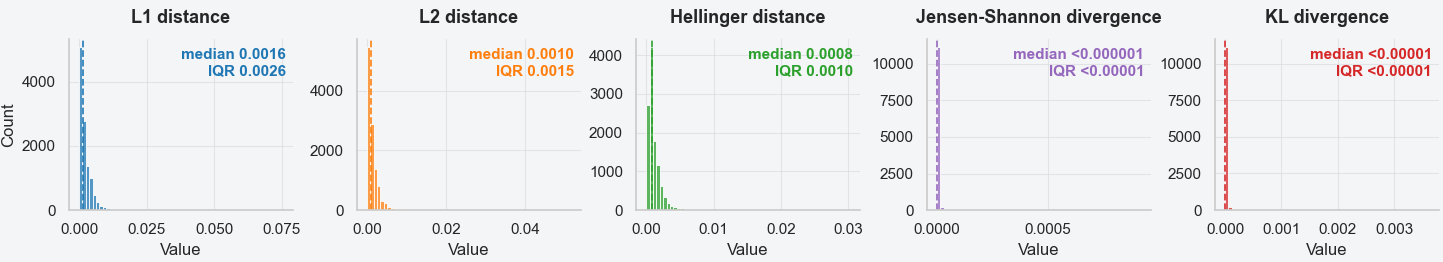}
  \caption{\textbf{In-distribution compression fidelity on GRCh38 (excluding Chromosome 1).}
  For Chr2--Chr22, X, and Y, we evaluate \ourmethod{}--Evo-2 7B against the baseline Evo-2 7B on $500$ random $1024$\,bp segments per chromosome.
  Histograms show the distribution of L1, L2, Hellinger, Jensen--Shannon, and $\mathrm{KL}$ across all positions; annotations mark the median and IQR.
  Most mass concentrates near zero, indicating high-fidelity compression.}
  \label{fig:GRCh38} 
\end{figure}

\subsection{In-Distribution Compression Fidelity}
\label{subsec:indist}
We first evaluate fidelity under \textbf{in-distribution} conditions, i.e., data aligned with the training distribution of the original DNA LLM (OpenGenome2).

For GRCh38 chromosomes Chr2--Chr22, X, and Y, we sample $500$ non-overlapping $1024$\,bp segments per chromosome.
As shown in \autoref{fig:GRCh38}, all five discrepancy measures are sharply concentrated near zero, indicating that the compressed model closely reproduces the baseline on the same inputs. 
Taking \textbf{L1 distance} as an example, the median value of $0.0016$ means that, at a typical position, the \emph{total} absolute change across the four next-base probabilities (A/C/G/T) is about $1.6\times10^{-3}$; equivalently, the \emph{average} per-nucleotide probability shifts by roughly $4\times10^{-4}$. 
In practical terms, such differences are negligible at the distribution level, consistent with near-lossless fidelity between the compressed and uncompressed models across held-out human chromosomes.

\begin{figure}[htbp]
  \centering
  \includegraphics[width=1\textwidth]{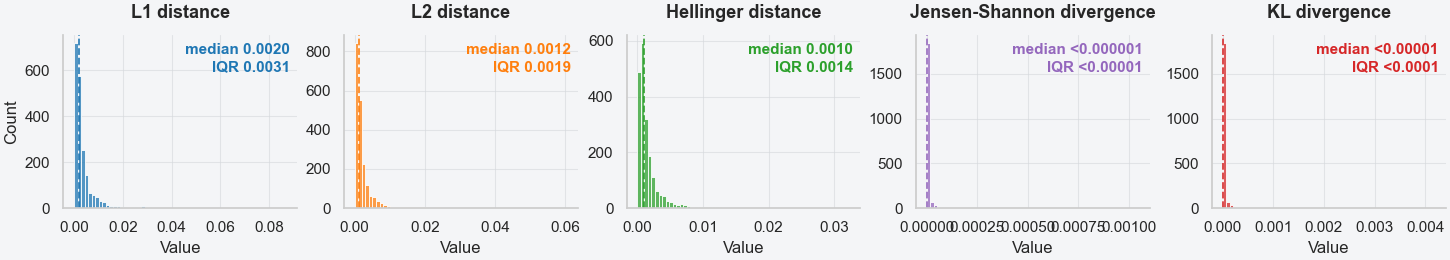}
  \caption{In-distribution compression fidelity on OpenGenome2.}
  \label{fig:OpenGenome2}
\end{figure}

We additionally sample $2{,}000$ random $1$\,kb segments from organisms represented in OpenGenome2, including both phage and eukaryotes. As shown in \autoref{fig:OpenGenome2}, all five discrepancy measures remain small, with medians slightly higher than on GRCh38. This is expected: the compression parameters of \ourmethod{} were tuned on GRCh38 (human) data, so the compressed model is marginally better adapted to human sequences. Overall, \ourmethod{} preserves the baseline behavior on \emph{in-distribution} sequences (with respect to the base DNA LLM's pretraining distribution).

\subsection{Out-of-Distribution Compression Fidelity}
\label{subsec:ood}
\begin{figure}[htbp]
  \centering
  \includegraphics[width=1\textwidth]{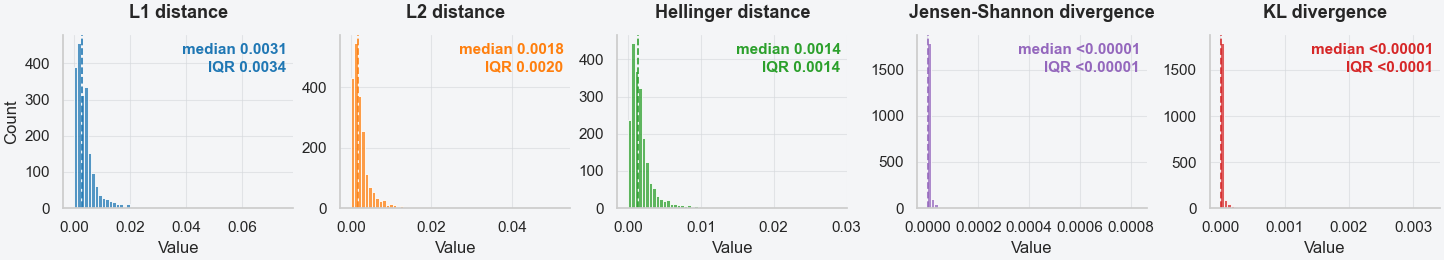}
  \caption{Out-of-distribution compression fidelity on MSL39 viruses sequences.}
  \label{fig:Virus}
\end{figure}

To test generalization beyond the training distribution, we evaluate on \textbf{MSL39 viral sequences}, which are not included in OpenGenome2.
From this corpus we sample $2,000$ random virus, each sampling a $1024$\,bp segment.
As shown in \autoref{fig:Virus}, the five metrics remain small in OOD conditions as well (e.g., median/IQR approximately $D_{\mathrm{L1}}=0.0020$ / $0.0031$, $D_{\mathrm{L2}}=0.0012$ / $0.0019$, $H=0.0010$ / $0.0014$; $\operatorname{JS}$ and $\operatorname{KL}$ medians $<10^{-5}$), indicating high-fidelity approximation on unseen genomic domains.

\subsection{Effect of Sequence Length and Hyperparameters on Fidelity}\label{subsec:length}
\begin{figure}[htbp]
  \centering
  \includegraphics[width=1\textwidth]{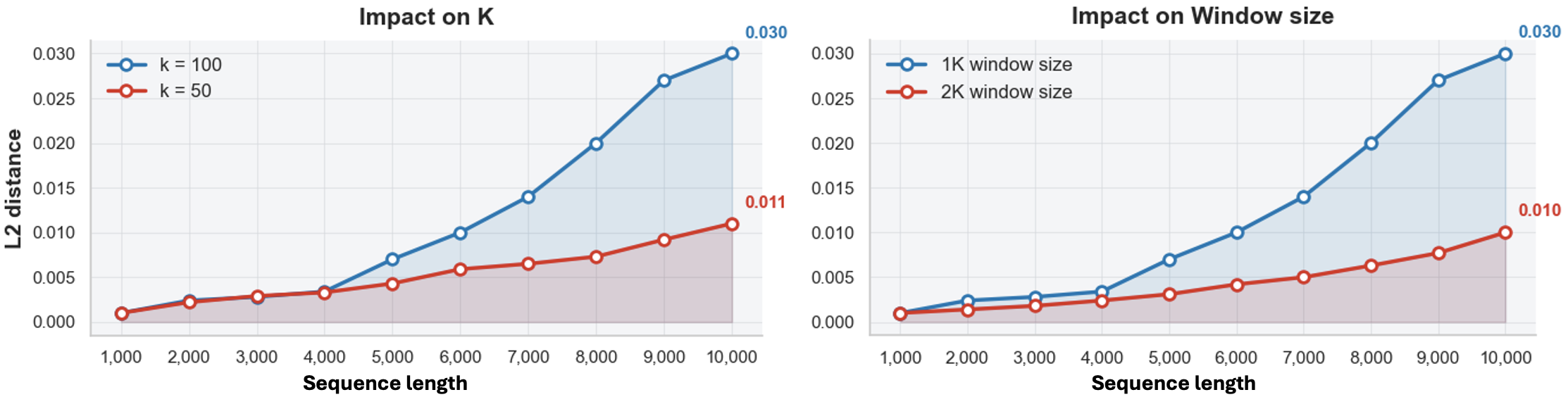}
  \caption{\textbf{Sequence length vs.\ L2 discrepancy, and effects of $W$ and $k$.}
  Left: median per-position L2 on GRCh38 for a model trained with $W{=}1024,k{=}100$ (blue) and a retrained model with $W{=}2048$ (red).
  Right: same protocol with $W{=}1024$ while changing $k$ from $100$ (blue) to $50$ (red).}
  \label{fig:impact_on}
\end{figure}

We examine how sequence length influences the approximation quality of the compressed model and how two key hyperparameters---window size ($W$) and $k$-granularity---modulate this effect.
Using the \ourmethod{}--Evo-2 7B trained on GRCh38 Chromosome 1 with $W{=}1024$ and $k{=}100$, we evaluate on GRCh38 by sampling, for each target length $L\in\{1\mathrm{k},2\mathrm{k},\ldots,10\mathrm{k}\}$, $100$ random non-overlapping segments.
For each set we compute the per-position L2 distance to the baseline Evo-2 7B and summarize it by the median across all tokens and sequences.
As shown in \autoref{fig:impact_on}, the discrepancy increases with length: longer inputs span more windows, so approximation errors accumulate across boundaries.
We then ablate the two hyperparameters.

(\emph{i}) Increasing the window size from $W{=}1024$ to $W{=}2048$ (same $k{=}100$ and training protocol) yields uniformly smaller errors, with the largest gains on the longest inputs (e.g., at $10$\,kbp, median L2 drops from $\approx 0.030$ to $\approx 0.010$).
Larger windows introduce fewer boundaries and preserve more within-window detail.
(\emph{ii}) Reducing the granularity from $k{=}100$ to $k{=}50$ (the same $W{=}1024$ and the same protocol) also improves fidelity across all lengths (e.g., at $10$\,kbp, $\approx 0.030 \rightarrow 0.011$) by inserting more \ourmethod{} tokens per window, thus providing finer summaries (\autoref{subsec:scalability}).

\subsection{Scalability: Near-Linear Complexity with Context Length}
\label{subsec:scalability}
\begin{figure}[htbp]
  \centering
  \includegraphics[width=0.7\textwidth]{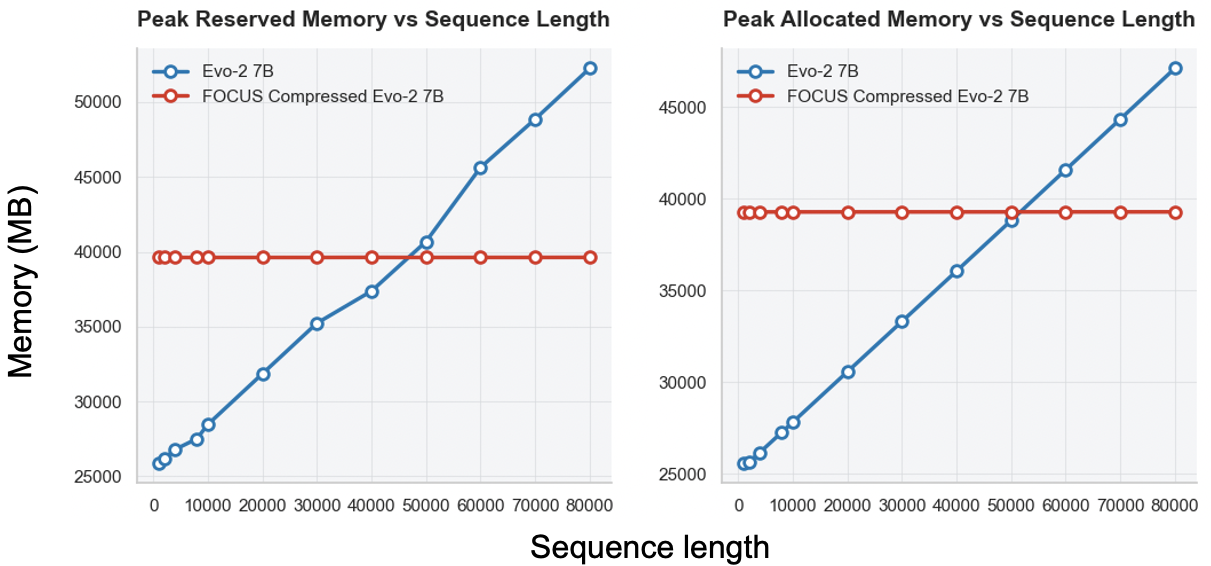}
  \caption{\textbf{Scalability of memory with context length.}
  Peak reserved (left) and peak allocated (right) GPU memory versus sequence length for Evo-2 7B (blue) and \ourmethod{}--Evo-2 7B (red).
  The baseline grows roughly linearly with context length, whereas \ourmethod{} is nearly constant (configuration $W{=}1024$, $k{=}100$), demonstrating the practical memory advantage for very long inputs.}
  \label{fig:Memory}
\end{figure}

Finally, we measure peak memory usage as a function of context length under identical hardware and batch settings.
\autoref{fig:Memory} compares the baseline Evo-2 7B with \ourmethod{}--Evo-2 7B.
The baseline exhibits near-linear growth in both peak \emph{reserved} and peak \emph{allocated} memory as the sequence length increases, while the \ourmethod{} curves remain essentially flat.
With $W{=}1024$ and $k{=}100$, \ourmethod{} retains only a small set of cross-window $\mathbf{KV}$ states, allowing single-GPU inference to scale to $\geq$80k tokens with almost constant memory. Note that, for sequences $\lesssim$50k tokens, \ourmethod{} shows higher peak memory than the baseline. This is expected: \ourmethod{} introduces a modest \emph{fixed} overhead (summary-token projections, per-window buffers, and retained summary $\mathbf{KV}$) that is largely independent of context length. 

  \vspace{-10 pt}
\section{Discussion \& Future work}
   \ourmethod\ compresses self-attention into $K$-mer--aware summary tokens and retains only their $\mathbf{KV}$ states across windows, delivering near-lossless fidelity while unlocking $\sim$100$\times$ longer contexts with near-linear memory. In practice, we observe a small fixed overhead at short lengths (from summary projections and per-window buffers), but as context grows the baseline's cache scales with raw tokens whereas \ourmethod's state scales with the number of summaries ($\mathcal{O}(L/K)$), yielding the flat memory curves central to genome-scale inference. Remaining limitations include fixed, globally chosen hyperparameters $(W,K)$ and mild error accumulation across many windows; both suggest principled adaptivity as the next step.

\noindent\textbf{Adaptive windowing.} A natural extension is to choose $W$ and $K$ \emph{on the fly} based on regulatory context. Smaller windows and finer $K$ near promoters, enhancers, splice sites, and other high-information loci can preserve detail; larger windows and coarser $K$ can accelerate traversal of quasi-neutral or repetitive regions. TAD-aware boundaries can align window transitions to chromatin domains to reduce cross-window leakage. 

\noindent\textbf{Multi-resolution and evaluation.} Beyond per-position fidelity, multi-resolution schedules---coarse summaries for navigation, fine summaries on demand---can further reduce cost, and retrieval-style memory could cache reusable summaries of recurrent repeats. We will evaluate adaptive \ourmethod\ on whole-genome pipelines with task-level metrics, ablations of adaptive policies, and robustness across species and assays. 

\newpage


\bibliographystyle{myrecomb}

\bibliography{mybib}

\newpage
\appendix 

\section{Formal Methodology}
\label{sec:Formal Methodology}

Given a DNA sequence $\mathbf{x}=(x_1,\dots,x_L)$, we partition it into consecutive \emph{windows} of fixed length $W$ (with $W\!\le\!N$, the base context size if applicable). Let the $w$-th window cover
\[
\mathcal{I}_w \;=\; [(w-1)W+1,\,wW]\cap[1,L], 
\qquad
M \;=\; \left\lceil \tfrac{L}{W}\right\rceil .
\]
Within each window, we \emph{insert} one \ourmethod\ token for every $k$ real bases, placed at a pre-defined offset (e.g., right boundary) relative to its associated $k$-mer. The number of \ourmethod\ tokens in window $w$ is
\[
B_w \;=\; \left\lceil \tfrac{|\mathcal{I}_w|}{k}\right\rceil ,
\]
and the set of all \ourmethod\ positions in window $w$ is denoted $\mathcal{B}_w$. Each \ourmethod\ token has its own learnable embedding and occupies a real sequence position; all attention is strictly causal.

\paragraph{Specialized attention and summarization.}
At a selected subset of decoder layers, we introduce a light-weight \emph{\ourmethod\ attention} unit to aggregate information into \ourmethod\ tokens. Let $H^{(\ell)}=[h^{(\ell)}_1,\dots,h^{(\ell)}_T]$ be the hidden states at layer~$\ell$ (including real and \ourmethod\ positions). For the $j$-th \ourmethod\ token in window $w$, indexed by $b_{w,j}$, denote its associated $k$-mer by $\mathcal{S}_{w,j}\subseteq \mathcal{I}_w$ (the $k$ real tokens immediately to its left). The layer-wise compression map is
\begin{equation}
h^{(\ell+1)}_{b_{w,j}}
\;=\;
f\!\big(\{\,h^{(\ell)}_t:\,t\in\mathcal{S}_{w,j}\,\};\;\Theta^{(\ell)}\big),
\label{eq:focus-compress}
\end{equation}
where $\Theta^{(\ell)}$ are learnable parameters. Concretely, $f(\cdot)$ is realized by masked multi-head attention with the \ourmethod\ token as the \emph{query}, and the visible set (keys/values) restricted by strict causality:
\begin{equation}
h^{(\ell+1)}_{b_{w,j}}
\;=\;
\mathrm{MHA}\!\Big(
Q=h^{(\ell)}_{b_{w,j}}W_Q^{(\ell)},\;
K=H^{(\ell)}_{\mathcal{V}_{b_{w,j}}}W_K^{(\ell)},\;
V=H^{(\ell)}_{\mathcal{V}_{b_{w,j}}}W_V^{(\ell)}
\Big),
\label{eq:focus-attn}
\end{equation}
with $\mathcal{V}_{b_{w,j}}\!\subseteq\!\mathcal{I}_w\cup(\cup_{w'<w}\mathcal{B}_{w'})$ containing only past positions within the current window and all past \ourmethod\ tokens from previous windows; future positions are never visible. Non-\ourmethod\ tokens may bypass this unit or be updated by standard decoder layers. Importantly, \ourmethod\ \emph{does not} explicitly compress hidden states beyond normal forward passes; compression is effected by which KV entries are retained across windows .

\paragraph{KV compression and budget.}
After finishing window $w$, we \emph{discard} KV for ordinary tokens in $\mathcal{I}_w$ and \emph{retain} only KV for $\mathcal{B}_w$ (plus a small boundary tail if required). After processing the first $L$ tokens, the retained KV count satisfies
\begin{equation}
R^{\text{kept}}_{\text{KV}}(L)
\;\le\;
\sum_{w=1}^{M}\Big\lceil\tfrac{W}{K}\Big\rceil \;+\; W
\;\approx\;
\tfrac{L}{K}+W,
\label{eq:kv-budget}
\end{equation}
compared to $R^{\text{full}}_{\text{KV}}(L)=L$ without compression. The average compression ratio is thus
\begin{equation}
\gamma
\;\triangleq\;
\frac{R^{\text{kept}}_{\text{KV}}(L)}{R^{\text{full}}_{\text{KV}}(L)}
\;\approx\;
\frac{1}{K}+\frac{W}{L},
\qquad
\text{so for } L\gg W:\;\gamma\approx \tfrac{1}{K}.
\label{eq:kv-ratio}
\end{equation}
Hence, choosing larger $k$ reduces the retained KV nearly in proportion to $1/K$.

\subsection{Formal Autoregressive Training}
\label{subsec:Compressed Autoregressive Fine-tuning}

\paragraph{Objective and visibility.}
Training follows next-token autoregressive learning while embedding a \emph{compression-aware visibility rule}. Let $\Phi$ denote backbone parameters and $\Theta$ the \ourmethod-related parameters (attention weights, \ourmethod\ embeddings, etc.). For any position $t$ (real or \ourmethod), define its visible context $\mathcal{C}(t)$ to include only past positions, and---across windows---\emph{only} the retained \ourmethod\ KV from earlier windows (ordinary-token KV from previous windows is \emph{not} visible). To force \ourmethod\ tokens to carry predictive content, we supervise them to predict the next \emph{real} base to their right. Let
\[
\pi(t)\;=\;\min\{\,u>t:\;u\text{ is a real-base position}\,\}.
\]
The cross-entropy loss is
\begin{equation}
\mathcal{L}(\Phi,\Theta)
\;=\;
-\sum_{t\in\mathcal{I}\cup\mathcal{B}}
\log p_{\Phi,\Theta}\!\big(x_{\pi(t)} \,\big|\, \mathcal{C}(t)\big),
\label{eq:loss}
\end{equation}
where $\mathcal{I}$ and $\mathcal{B}$ denote the sets of real-base and \ourmethod\ positions, respectively. For real-base $t$, \eqref{eq:loss} reduces to standard next-token prediction. For \ourmethod\ positions, \eqref{eq:loss} compels the \ourmethod\ representation to summarize sufficient information to predict the immediate next real base, thereby learning to \emph{encode} the $k$-mer (and longer-range) cues into a compact state.

\paragraph{Parameter updates and regularization.}
To introduce compression without eroding backbone competence, we typically \emph{freeze} most of $\Phi$ and optimize $\Theta$; light adapters (e.g., LoRA) or a small subset of high layers can be optionally unfrozen. For robustness, we can randomize compression configurations during training (e.g., enabling \ourmethod\ attention at different layers or slightly perturbing $k$), encouraging stable performance across compression strengths. All such choices fit within the single objective \eqref{eq:loss} without modifying the loss form.

\paragraph{Cross-window consistency.}
Two equivalent implementations enforce the same semantics: (i) \emph{sequential} windowed training, which processes windows in order and explicitly drops non-\ourmethod\ KV after each window; or (ii) \emph{masked-parallel} training on a longer segment, using attention masks to emulate that only earlier \ourmethod\ states are visible across window boundaries. In both cases, the definition of $\mathcal{C}(t)$ guarantees that past ordinary tokens are \emph{invisible} to later windows whereas past \ourmethod\ tokens remain \emph{visible}, aligning training-time visibility with inference-time memory.

\subsection{Inference-time Compression Strategy}
\label{sec:Inference-time Compression Strategy}
\paragraph{Streaming with controlled memory.}
At inference, we process windows $\mathcal{I}_1,\mathcal{I}_2,\dots,\mathcal{I}_M$ in order. Within each window, we insert \ourmethod\ tokens every $k$ real bases and apply the update in \eqref{eq:focus-attn}. After finishing window $w$, we \emph{discard} KV for ordinary tokens in $\mathcal{I}_w$ and \emph{retain} KV for $\mathcal{B}_w$ (plus a minimal boundary tail, if needed). The retained \ourmethod\ KV from prior windows is supplied as history when processing window $w\!+\!1$. This repeats until the entire sequence (or generated prefix) is covered. By \eqref{eq:kv-budget}--\eqref{eq:kv-ratio}, the retained KV grows approximately as $L/k$ (plus $W$), yielding a near-linear memory profile in $L$ with slope $1/k$.

\paragraph{Complexity and tunable trade-offs.}
Within each window, attention remains $O(W^2)$; across windows, the ``memory channel'' is carried by roughly $L/k$ \ourmethod\ tokens. Overall compute is thus $O(MW^2)$ with $M=\lceil L/W\rceil$, while \emph{memory} is governed by \eqref{eq:kv-budget} and scales $\approx L/k$. Larger $k$ strengthens compression (lower memory and latency) at the cost of coarser summaries; smaller $k$ improves fidelity but increases KV. A \emph{dynamic} policy is also possible: use stronger compression (larger $k$ or fewer enabled layers) for distant context and lighter compression near the current prediction point. For comparability, our evaluations adopt a fixed $(W,k)$ and layer configuration, but the interface to downstream tasks remains identical to the base decoder: given any-length DNA input, the model outputs next-base distributions, differing only in its internal maintenance of compact \ourmethod\ memory.

\section{Metrics}
\label{sec:Metrics}
\paragraph{(1) L1 distance}
\begin{equation}
D_{\mathrm{L1}}(\mathbf{p},\mathbf{q}) \;=\; \sum_{i=1}^{V} \left|p_i-q_i\right|.
\end{equation}

\paragraph{(2) L2 distance}
\begin{equation}
D_{\mathrm{L2}}(\mathbf{p},\mathbf{q}) \;=\;
\Bigl(\sum_{i=1}^{V} \bigl(p_i-q_i\bigr)^2 \Bigr)^{1/2}.
\end{equation}

\paragraph{(3) Hellinger distance}
\begin{equation}
H(\mathbf{p},\mathbf{q}) \;=\; \frac{1}{\sqrt{2}} \left\lVert \sqrt{\mathbf{p}}-\sqrt{\mathbf{q}} \right\rVert_2
\;=\; \sqrt{\,1-\sum_{i=1}^{V}\sqrt{p_i q_i}\,}.
\end{equation}

\paragraph{(4) Jensen--Shannon divergence}
\begin{equation}
\operatorname{JS}(\mathbf{p}\Vert\mathbf{q})
\;=\; \tfrac{1}{2}\operatorname{KL}\!\left(\mathbf{p}\middle\Vert\mathbf{m}\right)
+\tfrac{1}{2}\operatorname{KL}\!\left(\mathbf{q}\middle\Vert\mathbf{m}\right),
\qquad \mathbf{m}=\tfrac{1}{2}\,(\mathbf{p}+\mathbf{q}),
\end{equation}
with $\operatorname{KL}(\mathbf{a}\Vert\mathbf{b})=\sum_i a_i\log\!\frac{a_i}{b_i}$ (natural log; units in nats).

\paragraph{(5) KL divergence (base $\Vert$ \ourmethod{})}
\begin{equation}
\operatorname{KL}_{\mathrm{base}\Vert\ourmethod{}}(\mathbf{p}\Vert\mathbf{q})
\;=\; \sum_{i=1}^{V} p_i \log\!\frac{p_i}{q_i}.
\end{equation}

For numerical stability we apply probability clipping at $10^{-12}$ and re-normalization.
A sequence-level score can also be defined as $\overline{D}(x)=\frac{1}{T}\sum_{t=1}^{T} D(\mathbf{p}_t,\mathbf{q}_t)$; the figures visualize the overall distribution of per-position metrics across all $(t,x)$ together with the median and IQR.

\section{FOCUS hyper-parameters}
\label{sec:FOCUS hyper-parameters}
\begingroup
\setlength{\LTleft}{0pt}\setlength{\LTright}{0pt}
\setlength{\tabcolsep}{3.5pt}
\renewcommand{\arraystretch}{1.05}
\footnotesize

\newcolumntype{P}[1]{>{\raggedright\arraybackslash}p{#1}}

\begin{longtable}{@{}P{0.30\textwidth}P{0.22\textwidth}P{0.48\textwidth}@{}}
\caption{Full configuration.}
\label{tab:full_config}\\
\toprule
\textbf{Parameter (symbol)} & \textbf{Value} & \textbf{Meaning} \\
\midrule
\endfirsthead
\toprule
\textbf{Parameter (symbol)} & \textbf{Value} & \textbf{Meaning} \\
\midrule
\endhead
\midrule \multicolumn{3}{r}{\emph{(continued)}}\\ \midrule
\endfoot
\bottomrule
\endlastfoot

Backend & BioNeMo & Adapter on NeMo/Megatron. \\
Backbone $\mathcal{M}$ & Evo-2 7B  & Base LLM; frozen. \\
Window size $W$ & 1024 & Block length. \\
Cadence $k$ & 100 & Insert one FOCUS per $k$ bases. \\
\#FOCUS per window $b$ & 1 & FOCUS tokens per window. \\
Condense factor $r$ & $\sim$100$\times$ & Approx KV compression. \\
Freeze base & \texttt{true} & Train adapter + FOCUS embedding. \\
Max position $L_{\max}$ & 131{,}072 & Long-context range. \\
Special FOCUS token & reserved & Tokenizer marker. \\
Tokenizer & byte-level & Byte-ID tokens. \\
Random ratio sampling & \texttt{false} & Fixed cadence ($[100,100]$). \\[2pt]

Epochs & 1 & One pass on Chr1. \\
Learning rate & $1\!\times\!10^{-4}$ & Adapter LR. \\
Weight decay & 0.01 & Regularization. \\
Warmup steps & 200 & LR warmup. \\
Batch size (per device) & 1 & Micro-batch. \\
Grad.\ accumulation & 4 & Effective batch $=4$. \\
Mixed precision & \texttt{bf16} & Lower mem; faster. \\
Seed & 42 & Reproducibility. \\
Dataset loader & JSONL (Chr1), field \texttt{text} & One DNA chunk per record. \\[2pt]

Max new tokens & 1024 & Generation cap. \\
Temperature & 0.8 & Sampling temperature. \\
Top-$p$ & 0.95 & Nucleus sampling. \\
Repetition penalty & 1.0 & Disabled. \\
Sliding-window stride & 512 & Optional rolling eval. \\[2pt]

Success: memory\_drop\_min & 0.30 & Min GPU memory reduction. \\
Success: quality\_drop\_max & 0.05 & Max metric drop vs baseline. \\
Success: speedup\_min & 0.00 & No speedup required. \\[2pt]

Metric: eval task & Perplexity (Chr1 val, $\le$512) & Primary quality metric. \\

Platform: seq\_length & 512 & Internal micro length. \\
Platform: FP8 & \texttt{false} & FP8 off. \\
Platform: TE kernels & \texttt{use\_te=true} & Transformer Engine on. \\[2pt]

\end{longtable}
\endgroup
\end{document}